\documentclass[11pt]{article}
\usepackage{epsfig} 
\usepackage{graphics}
\newcommand{\sect}[1]{ \section{#1} \setcounter{equation}{0} } 

\newcommand{\half}{\mbox{\small{$\frac{1}{2}$}}}

\newcommand{\MSbar}{\overline{\mbox{MS}}} 
\newcommand{\Nc}{N_{\!c}}
\newcommand{\Nf}{N_{\!f}}
\newcommand{\bare}{\mbox{\footnotesize{o}}}

\begin{document}

\title{Renormalization of scalar field theories in rational spacetime
dimensions}
\author{J.A. Gracey, \\ Theoretical Physics Division, \\ 
Department of Mathematical Sciences, \\ University of Liverpool, \\ P.O. Box 
147, \\ Liverpool, \\ L69 3BX, \\ United Kingdom.} 
\date{}
\maketitle 

\vspace{5cm} 
\noindent 
{\bf Abstract.} We renormalize various scalar field theories with a $\phi^n$ 
self interaction such as $n$~$=$~$5$, $7$ and $9$ in their respective critical
dimensions which are non-integer. The renormalization group functions for the
$O(N)$ symmetric extensions are also computed. 

\vspace{-15.0cm}
\hspace{13cm}
{\bf LTH 1130}

\newpage 

\sect{Introduction.}

Scalar quantum field theories have provided an excellent laboratory for many
years to explore and test ideas in physical problems. For instance, the
development of Wilson's renormalization group and its application to theories
defined close to an integer spacetime dimension led to the concept of the
Wilson-Fisher fixed point, \cite{1,2,3,4,5}. Being defined as a non-trivial 
zero of the $\beta$-functions in spacetime dimensions 
$d$~$=$~$D_n$~$-$~$2\epsilon$, where $D_n$ is the critical dimension which will
be defined later for a scalar theory and is not necessarily an integer, meant 
that the renormalization group functions in the neighbourhood of a fixed point 
provided information on phase transitions in nature, \cite{1,2,5}. A widely 
studied example is that of the Ising model which can be described by scalar 
$\phi^4$ theory. In this case $D_n$~$=$~$4$, while other theories such as 
scalar $\phi^3$ have $D_n$~$=$~$6$ and this potential underpins the properties 
of phase transitions for other phenomena. For example, the Lee-Yang singularity
problem can be accessed via $\phi^3$ theory. Recently there has been renewed 
interest in examining scalar $\phi^n$ theories for $n$~$\geq$~$4$. Early work 
on such higher order potentials included articles on $\phi^6$ theory, 
\cite{6,7,8,9,10,11,12}, and $\phi^{2r}$ potentials for integer $r$, 
\cite{13,14}. However, more recently scalar $\phi^5$ theory was studied in 
\cite{15} using the functional renormalization group method as well as via 
critical exponents in \cite{16,17,18}. The motivation was to develop and 
investigate the continuum quantum field theory for the Blume-Capel universality
class which is the next after the Ising and Lee-Yang classes. Equally as phase 
transitions have scale and conformal symmetry, the formalism of conformal field
theory has been used to calculate exponents and central charges associated with
the correlation functions of various operators or currents, \cite{17,18,19,20}.
These were then used to inform the structure of the perturbative 
renormalization group functions. One aim is partly to continue building up the 
formalism associated with $d$-dimensional conformal field theory as well as to 
have a powerful tool to make predictions for new universality classes such as 
the Blume-Capel one.  

Given this resurgence of interest in scalar field theories with higher order
potentials there is a clear need to complement the conformal field theory
approach with explicit perturbative computations. This is the purpose of this
article. Aside from $\phi^4$ and $\phi^3$ theories whose renormalization group
functions are known to high order, 
\cite{21,22,23,24,25,26,27,28,29,30,31,32,33,34}, only $\phi^6$ theory has been 
renormalized to any depth, \cite{8,9,12}. Some leading order results are 
available for $\phi^{2r}$ theories for integer $r$~$\geq$~$4$ but those for 
potentials with odd powers except $3$ are virtually unknown. Therefore we will 
consider the $\phi^5$, $\phi^7$ and $\phi^9$ scalar field theories and
determine the anomalous dimensions and $\beta$-functions as well as those for 
$\phi^8$ for a reason which will become apparent later. While this is simple to
state it is worth observing that these scalar theories are not renormalizable 
in integer dimensions. By contrast their critical dimension is rational. 
Although this is clearly not a value for a physical spacetime such models 
should in principle provide more accurate predictions for physical phase 
transitions. As has been noted for example in \cite{15} if the critical 
dimension is close to an integer then the use of $d$~$=$~$D_n$~$-$~$2\epsilon$ 
means that choosing a small value of $\epsilon$ in the $\epsilon$-expansion of 
the first few terms of a critical exponent should yield accurate exponent
estimates in that integer dimension. This is in contrast with the use of the 
$\epsilon$-expansion in $\phi^4$ theory where $d$~$=$~$4$~$-$~$2\epsilon$ and 
$d$ has to be $3$ for Ising model predictions which requires the relatively 
high value of $\epsilon$~$=$~$\half$. In this context $\phi^4$ theory can also 
be used to access other physical phenomena if it is endowed with an $O(N)$ 
symmetry. Then, for instance, $N$~$=$~$2$ describes superfluidity while 
$N$~$=$~$3$ corresponds to the Heisenberg magnet. Therefore in this spirit we 
will extend the odd higher potentials to include an $O(N)$ symmetry and compute
the corresponding renormalization group functions. This leads to an interesting
prospect which may connect the $O(N)$ $\phi^5$ and $O(N)$ $\phi^8$ theories
with potentially a generalization for higher order potentials. Such connections
should be established by explicit computations. For the widely known 
ultraviolet completion of $O(N)$ $\phi^4$ theory in four dimensions to $O(N)$ 
$\phi^3$ theory in six dimensions, \cite{33,35}, this has been put on a 
concrete foundation via higher order perturbative computations and the large 
$N$ expansion. Equally the $d$-dimensional conformal field theory formalism is 
in a position to address the same connection in principle. Hence it ought to be
a crucial tool for the new connections that we suggest are apparent here in the 
higher order potentials. Therefore providing renormalization group functions in
this article will inform that debate.

The article is organized as follows. The following section reviews the 
background to scalar theories with $\phi^n$ potentials. Results for the scalar
theories with odd potentials are given in section $3$ while the corresponding 
results when an $O(N)$ symmetry is present are given in the subsequent section 
together with the potential connections between theories with odd and even
potential terms. Concluding remarks are provided in section $5$. 

\sect{Background.}

Our starting point is the general scalar field theory given by the Lagrangian
\begin{equation}
L^{(n)} ~=~ \frac{1}{2} \partial_\mu \phi_{\bare} \partial^\mu \phi_{\bare} ~+~
\frac{g_{\bare}}{n!} \phi_{\bare}^n 
\label{lagn}
\end{equation}
where $n$~$\geq$~$3$ is an integer and for the moment we do not endow the 
theory with a symmetry group. Bare fields and variables will be denoted by the 
subscript ${}_{\bare}$ throughout and these are related to the corresponding 
renormalized quantities via the renormalization constants such as
\begin{equation}
\phi_{\bare} ~=~ \sqrt{Z_\phi} \phi ~.
\end{equation} 
We will give the coupling constant renormalization constant later. The critical
dimension $D_n$ of the theory where it is renormalizable is found by examining 
the canonical dimensions of the field $\phi_{\bare}$ and the coupling constant 
$g_{\bare}$ in $d$-dimensions. Ensuring that the action is dimensionless 
implies that the canonical dimension of $\phi_{\bare}$ is
\begin{equation}
[ \phi_{\bare} ] ~=~ \frac{1}{2} d ~-~ 1 ~.
\end{equation} 
From the interaction the canonical dimension of the bare coupling 
constant is therefore
\begin{equation}
[ g_{\bare} ] ~=~ d ~+~ n ~-~ \frac{1}{2} d n ~.
\end{equation} 
The critical dimension where the field theory is purely renormalizable is then
the spacetime dimension $D_n$ which is the solution of $[g_{\bare}]$~$=$~$0$. 
In other words 
\begin{equation}
D_n ~=~ \frac{2n}{(n-2)} 
\label{critd}
\end{equation}
For $n$~$=$~$3$ and $4$ we retrieve the usual integer critical dimensions of 
$6$ and $4$ respectively for a scalar cubic and quartic interaction. For the 
next two values of $n$ we have $D_5$~$=$~$\frac{10}{3}$ and $D_6$~$=$~$3$. As 
noted in \cite{15} $D_n$ is a monotonically decreasing function with
$D_n$~$\to$~$2$ in the limit as $n$~$\to$~$\infty$. So there are no more 
integer critical dimensions for $n$~$\geq$~$7$. For instance, 
$D_7$~$=$~$\frac{14}{5}$ and $D_8$~$=$~$\frac{8}{3}$. As the renormalization 
group functions for the theories with integer critical dimensions have been 
extensively studied, \cite{8,9,12,21,22,23,24,25,26,27,28,29,30,31,32,33,34}, 
the next step in studying scalar theories of the form (\ref{lagn}) are those 
with non-integer critical dimensions which is our main aim. 

In order to be able to carry this out we need to make certain reasonable
assumptions. For instance, we take the point of view that there is nothing
special about the critical dimension being non-integer and have defined bare 
and renormalized fields and parameters in the usual manner. As we will be 
deriving the renormalization group functions for renormalizable theories then
for the set of Lagrangians given in (\ref{lagn}) we do not need more 
renormalization constants than are available from the pure rescaling of the 
quantities present in (\ref{lagn}). The main obstacle to be overcome is the 
extraction of the divergences from the $2$- and $n$-point functions of each 
Lagrangian. In integer critical dimensional theories one has regularizations 
such as cutoff and dimensional regularization available, for example. In these 
regularizations the ultraviolet divergence arises in the integration over the 
radial components of the interal loop momenta and not the angular variables. 
The situation for non-integer critical dimensions is clearly the same. However,
what is difficult to handle immediately in cutoff regularization is the 
definition of angular integrals. This effectively results in using dimensional 
regularization as the method to extract the ultraviolet divergences. The 
regularization is introduced by analytically extending the spacetime dimension 
$d$ to
\begin{equation}
d ~=~ D_n ~-~ 2 \epsilon ~.
\label{ddef}
\end{equation}
This has the advantage that we can simply apply the well-established techniques
to evaluate dimensionally regularized Feynman integrals. With (\ref{ddef}) the 
dimension of the bare coupling in the dimensionally regularized theory is
\begin{equation}
[ g_{\bare} ] ~=~ ( n - 2 ) \epsilon 
\end{equation} 
which means that to retain a dimensionless renormalized coupling constant with
this regularization we take
\begin{equation}
g_{\bare} ~=~ \mu^{(n-2)\epsilon} Z_g g
\end{equation}
which defines $Z_g$ as the coupling constant renormalization constant. The 
arbitrary mass scale $\mu$ is introduced to balance the dimensions. Once the 
regularization and renormalization constants have been introduced we can define
our renormalization group functions. The $\beta$-function and field anomalous
dimension are given by
\begin{equation}
\beta(g) ~=~ \mu \frac{dg}{d\mu} ~~~,~~~
\gamma_\phi(g) ~=~ \mu \frac{\partial ~}{\partial \mu} \ln Z_\phi ~.
\end{equation}
Since the bare coupling has no $\mu$ dependence then in practical terms one 
finds the terms in the perturbative expansion of the $\beta$-function by 
iteratively solving
\begin{equation}
(n-2) \epsilon g Z_g ~+~ \mu \frac{d~}{d \mu} \left( g Z_g \right) ~=~ 0 ~.
\label{betaderiv}
\end{equation}
We have not applied the product and chain rules to the second term of
(\ref{betaderiv}) as later we will be considering theories with more than one 
coupling constant. In that case the differentiaton involves the 
$\beta$-functions of all the coupling constants. However, it is instructive to 
note that for all $n$~$\geq$~$3$ the leading solution of (\ref{betaderiv}) 
gives
\begin{equation}
\beta(g) ~=~ -~ (n-2) \epsilon g ~+~ O(g^2)
\end{equation}
or  
\begin{equation}
\beta(g) ~=~ \half (n-2) ( d - D_n ) g ~+~ O(g^2) ~.
\end{equation}
We have not included the higher order terms from the loop calculations as the
dependence of $g$ differs depending on whether $n$ is even or odd as will be
apparent from the explicit expressions given later.

{\begin{table}[ht]
\begin{center}
\begin{tabular}{|c||c|c|c|c||c|}
\hline
$n$ & $2$-point LO & $2$-point NLO & vertex LO & vertex NLO & $L_v$ \\
\hline
$3$ & $1$ & $2$ & $1$ & $7$ & $1$ \\
$4$ & $0$ & $1$ & $3$ & $9$ & $1$ \\
$5$ & $1$ & --- & $35$ & --- & $3$ \\
$6$ & $0$ & $1$ & $10$ & $115$ & $4$ \\
$7$ & $1$ & --- & $357$ & --- & $5$ \\
$8$ & $0$ & $1$ & $35$ & $1085$ & $6$ \\
$9$ & $1$ & --- & $3271$ & --- & $7$ \\
\hline
\end{tabular}
\end{center}
\begin{center}
{Table 1. Number of graphs computed at various orders for $n$ given in
(\ref{lagn}).}
\end{center}
\end{table}}

{\begin{figure}[hb]
\begin{center}
\includegraphics[width=12cm,height=1.7cm]{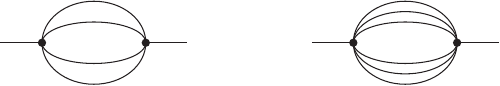}
\end{center}
\caption{Leading order topologies for $2$-point functions in $\phi^5$ and 
$\phi^7$ scalar theories.} 
\label{2ptlo57}
\end{figure}}

It is straightforward to evaluate the core Feynman integrals contributing to 
the renormalization of the wave function, coupling constant and mass operator 
and note that we have set up an automatic computation to handle the large 
number of graphs that arise with high order potentials. The Feynman graphs to 
be computed for the wave function and coupling constant renormalizations are 
generated using the {\sc Qgraf} package, \cite{36}. The numbers of graphs which
were evaluated for each Green's function for $n$~$\geq$~$5$ are given in Table 
$1$ where LO and NLO mean leading order and next to leading order respectively.
The final column gives $L_v$ which is the number of loops in the leading order 
vertex Green's functions. The actual independent topologies, rather than all 
the diagrams, for various theories are illustrated in various Figures 
throughout. The structure of the $2$-point graphs computed for $\phi^5$ and 
$\phi^7$ is shown in Figure \ref{2ptlo57}. That for $\phi^9$ theory is obtained
by adding additional internal propagators joining the vertices with external 
legs. The remaining Figures \ref{verlo5} to \ref{vernlo8} show the topologies 
of the various vertex functions for the odd potentials as well as $2$-point and
vertex functions for low order even potentials for comparison. In Table $1$ the
data given for $n$~$=$~$3$ and $4$ are for reference and comparison purposes 
only as the renormalization group functions for these theories are already well
established to very high loop order. Once the {\sc Qgraf} output is generated 
for the Green's functions of a theory, it can be adapted for the application of
our integration algorithm to each individual graph. Once these have all been 
evaluated then they are summed. Essential in this process is the symbolic 
manipulation language {\sc Form}, \cite{37,38}. The final step is the summation
of the individual divergences and the renormalization. The latter is effected 
by the method of \cite{39} whereby we compute all the integrals in terms of the
bare parameters. Then the counterterms are introduced automatically by 
rescaling to the corresponding renormalized variables. The constant of 
proportionality is the respective renormalization constant. All our final 
renormalization group functions will be in the $\MSbar$ scheme.

{\begin{figure}[ht]
\begin{center}
\includegraphics[width=12cm,height=4cm]{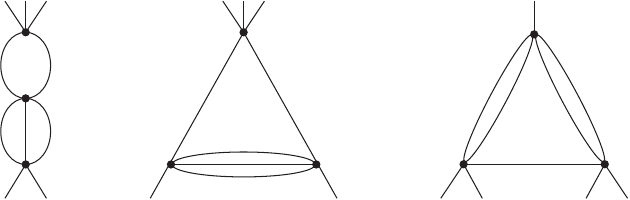}
\end{center}
\caption{Leading order topologies for $5$-point function in scalar $\phi^5$ 
theory.} 
\label{verlo5}
\end{figure}}

{\begin{figure}[ht]
\begin{center}
\includegraphics[width=14cm,height=10cm]{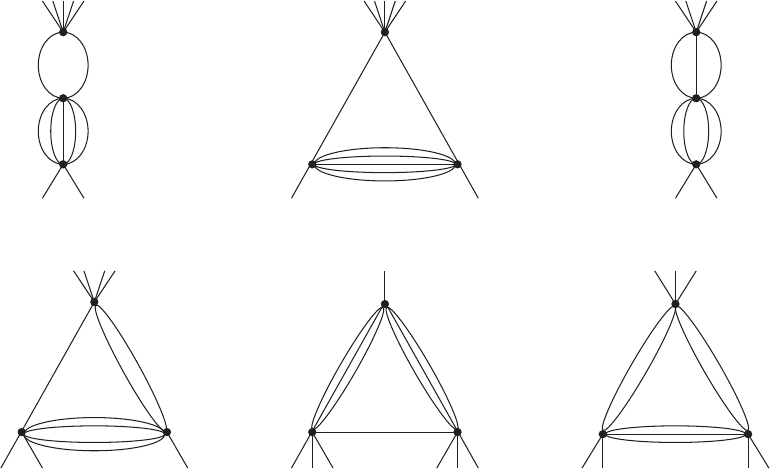}
\end{center}
\caption{Leading order topologies for $7$-point functions in scalar $\phi^7$ 
theory.} 
\label{verlo7}
\end{figure}}

\sect{Results.}

Having outlined the computation methodology we now present the results for the
various theories. First, for the case of $n$~$=$~$5$ we have 
\begin{eqnarray}
\beta^{\phi^5}(g) &=& -~ 3 \epsilon g ~+~
\frac{1377}{16} \Gamma^3 \left( \frac{2}{3} \right) g^3 ~+~ O(g^5) \nonumber \\
\gamma^{\phi^5}_\phi(g) &=& 
-~ \frac{3}{80} \Gamma^3 \left( \frac{2}{3} \right) g^2 ~+~ O(g^4) \nonumber \\
\gamma^{\phi^5}_{\cal O}(g) &=& 
\frac{63}{80} \Gamma^3 \left( \frac{2}{3} \right) g^2 ~+~ O(g^4) ~.
\end{eqnarray}
For this and some other cases we will include the anomalous dimension of the 
mass operator
\begin{equation}
{\cal O} ~=~ \frac{1}{2} \phi^2 ~.
\end{equation} 
Its renormalization is carried out by inserting the operator in a $2$-point 
function. Also the graphs which contribute are generated by {\sc Qgraf} but we 
have not illustrated these graphically. Instead they can be deduced from the 
vertex topologies as they can be derived from graphs where there are only two 
internal propagators connecting with an external vertex. Replacing that vertex 
by the operator ${\cal O}$ gives a contributing topology to the $2$-point 
function for the operator renormalization. As an application of these 
renormalization group functions we have evaluated the critical exponent 
$\sigma$ which was estimated in \cite{15} using functional renormalization 
group methods. The exponent is defined by the hyperscaling relation
\begin{equation}
\sigma ~=~ \frac{(d-2+2\eta)}{(d+2-2\eta)}
\end{equation}
where $d$~$=$~$\frac{10}{3}$~$-$~$2\epsilon$, 
$\eta$~$=$~$\gamma^{\phi^5}_\phi(g_c)$ and $g_c$ is the non-zero critical 
coupling constant given by the solution of $\beta^{\phi^5}(g_c)$~$=$~$0$. It 
corresponds to a Wilson-Fisher fixed point. We find 
\begin{equation}
\sigma ~=~ \frac{1}{4} ~-~ \frac{115}{408} \epsilon ~+~ O(\epsilon^2)
\end{equation}
or
\begin{equation}
\sigma ~=~ 0.250000 ~-~ 0.281863 \epsilon ~+~ O(\epsilon^2)
\end{equation}
numerically. As has been widely noted since the critical dimension of this 
theory is close to an integer dimension then the convergence of the $\epsilon$ 
expansion ought to be faster than say using the $\epsilon$ expansion of 
$\phi^4$ theory to extract exponent estimates in three dimensions. For 
$n$~$=$~$5$ we find 
\begin{equation}
\left. \sigma \right|_{d=3} ~=~ 0.203023 
\end{equation}
which is in agreement with \cite{17}. The value is not unreasonable for a
{\em leading} order computation when compared to the value of $0.198$ for 
$\sigma$ using functional renormalization group methods of \cite{15}. Also we 
have computed the exponent $1/\nu$ from $\gamma^{\phi^5}_{\cal O}(g)$ and note 
that it agrees with \cite{17}.

{\begin{figure}[ht]
\begin{center}
\includegraphics[width=16cm,height=15cm]{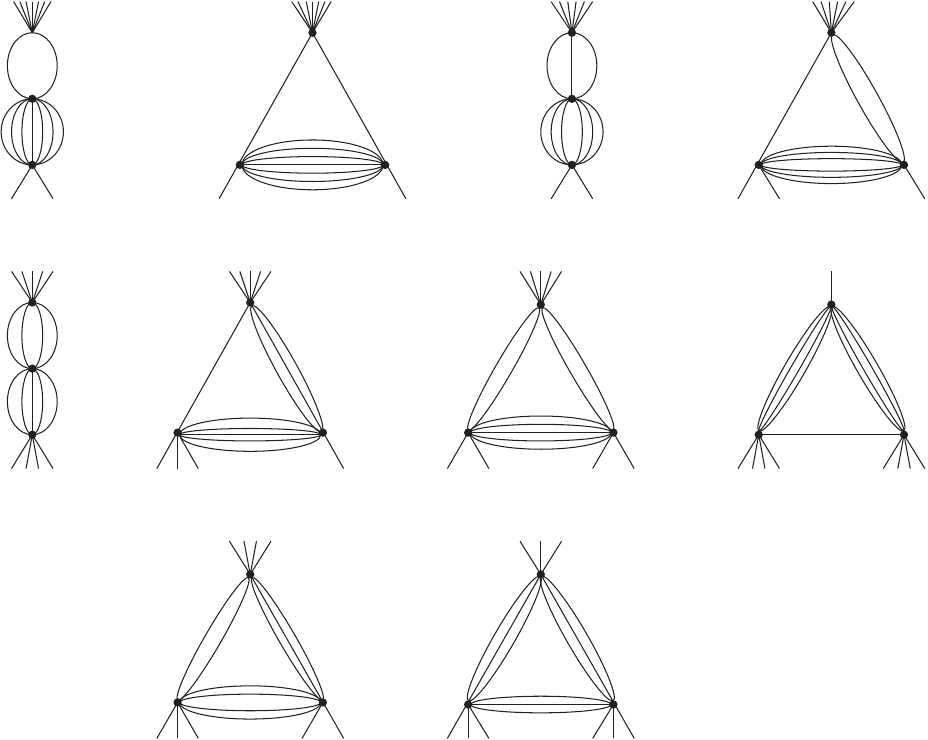}
\end{center}
\caption{Next to leading order topologies for $9$-point function in $\phi^9$ 
theory.} 
\label{verlo9}
\end{figure}}

For the next two odd power potentials we find the following sets of
renormalization group functions  
\begin{eqnarray}
\beta^{\phi^7}(g) &=& -~ 5 \epsilon g ~+~
\frac{25}{288} \Gamma^5 \left( \frac{2}{5} \right)
\left[ 1733 + 1260 \frac{\Gamma^2 \left( \frac{3}{5} \right)
\Gamma \left( \frac{2}{5} \right)}{\Gamma^2 \left( \frac{4}{5} \right) }
\right] g^3 ~+~ O(g^5) \nonumber \\
\gamma_\phi^{\phi^7}(g) &=&
-~ \frac{5}{2016} \Gamma^5 \left( \frac{2}{5} \right) g^2 ~+~ O(g^4) 
\nonumber \\
\gamma_{\cal O}^{\phi^7}(g) &=&
\frac{25}{672} \Gamma^5 \left( \frac{2}{5} \right) g^2 ~+~ O(g^4)
\end{eqnarray}
and
\begin{eqnarray}
\beta^{\phi^9}(g) &=& -~ 7 \epsilon g \nonumber \\
&& +~ \left[ -~ 317520 \Gamma^4 \left( \frac{6}{7} \right) 
\Gamma^2 \left( \frac{5}{7} \right) \Gamma \left( \frac{2}{7} \right)
+ 364899 \Gamma^3 \left( \frac{6}{7} \right) 
\Gamma^2 \left( \frac{4}{7} \right) \Gamma \left( \frac{3}{7} \right) 
\right. \nonumber \\
&& \left. ~~~~
+ 756000 \Gamma^2 \left( \frac{6}{7} \right) 
\Gamma \left( \frac{5}{7} \right) \Gamma \left( \frac{4}{7} \right) 
\Gamma^2 \left( \frac{3}{7} \right) \Gamma \left( \frac{2}{7} \right) 
+ 32000 \Gamma^2 \left( \frac{4}{7} \right) 
\Gamma^4 \left( \frac{3}{7} \right) \Gamma \left( \frac{2}{7} \right) \right]
\nonumber \\
&& ~~~~ \times \frac{49 \Gamma^7 \left( \frac{2}{7} \right) g^3} 
{172800 \Gamma^3 \left( \frac{6}{7} \right) \Gamma^2 \left( \frac{4}{7} \right)
\Gamma \left( \frac{3}{7} \right) } ~+~ O(g^5)
\nonumber \\
\gamma^{\phi^9}_\phi(g) &=&
-~ \frac{7}{103680} \Gamma^7 \left( \frac{2}{7} \right) g^2 ~+~ O(g^4) 
\end{eqnarray}
where $D_9$~$=$~$\frac{18}{7}$. In comparison with $\beta^{\phi^5}(g)$ both 
$\beta$-functions have a new feature in that there are two distinct terms in 
contrast to the one of the $n$~$=$~$5$ theory. By this we mean two independent 
combinations of $\Gamma$-functions. 

It transpires, however, that this first occurs in the $n$~$=$~$6$ theory since 
\begin{eqnarray}
\beta^{\phi^6}(g) &=& -~ 4 \epsilon g ~+~ \frac{20\pi}{3} g^2 ~-~ 
\left[ 225 \pi^2 + 2248 \right] \frac{\pi^2 g^3}{30} ~+~ O(g^5) \nonumber \\
\gamma^{\phi^6}_\phi(g) &=& \frac{\pi^2}{45} g^2 ~-~ \frac{4\pi^3}{81} g^3 ~+~ 
O(g^4) ~.
\end{eqnarray}
These were computed in \cite{8,9} and extended to the next order in \cite{12}. 
We evaluated them here as a check on our computation method before applying it 
to the odd power potentials. While there was a mismatch in the one loop terms 
of the $n$~$=$~$7$ theory it occurs first at two loops for $n$~$=$~$6$. 
Therefore we expect that the first occurrence of the mismatch for $n$~$=$~$5$ 
will be at two loops. For $n$~$=$~$8$ the $\beta$-function has a similar 
structure to $n$~$=$~$6$ since 
\begin{eqnarray}
\beta^{\phi^8}(g) &=& -~ 6 \epsilon g ~+~
\frac{70 \pi^3 \sqrt{3}}{9 \Gamma^3 \left( \frac{2}{3} \right)} g^2 
\nonumber \\
&& +~ \left[ 297675 \pi \sqrt{3} \Gamma^3 \left( \frac{2}{3} \right)
- 313600 \pi^3 \sqrt{3} - 893025 \ln 3 \, \Gamma^3 \left( \frac{2}{3} \right)
\right. \nonumber \\
&& ~~~~~ \left.
-~ 3082536 \Gamma^3 \left( \frac{2}{3} \right) \right] 
\frac{4\pi^6 g^3}{25515 \Gamma^9 \left( \frac{2}{3} \right)} ~+~ O(g^4)
\nonumber \\
\gamma^{\phi^8}(g) &=& 
\frac{2\pi^6}{945 \Gamma^6 \left( \frac{2}{3} \right)} g^2 ~-~ 
\frac{\pi^9 \sqrt{3}}{162 \Gamma^9 \left( \frac{2}{3} \right)} g^3 ~+~ 
O(g^4) ~.
\end{eqnarray} 
One of the reasons for highlighting this aspect of the renormalization group 
function numerology is that the differences reflect the underlying topologies 
of each vertex function as well as certain properties. For instance one can 
define a weighted sum for any of the products of $\Gamma$-functions appearing 
in the $\beta$-functions by summing the products of the $\Gamma$-function 
arguments with the power. Examining $\beta^{\phi^9}(g)$ the respective index of
each of the four terms are $3$, $2$, $2$ and $1$ where we count the 
contribution from a denominator $\Gamma$-function in the sum as negative. For 
$n$~$=$~$5$ we have $2$ for both terms in the $\beta$-function. For the even 
dimensional cases the various powers of $\pi$, $\ln 3$ and $\sqrt{3}$ would 
first have to be re-expressed in terms of $\Gamma$-functions. For the 
well-studied cases of $n$~$=$~$3$ and $4$ the same aspect of weighting is 
present. The difference is that the $\Gamma$-functions are already hidden in 
the corresponding expression as they will have arisen with integer arguments. 
Although drawing attention to this particular weighting or property of the 
renormalization group functions in rational dimensions may appear to be a 
quirk, it is in fact a guide to the expectations of the series of various sums 
which can appear at higher loop order. For instance the corrections to the 
$n$~$=$~$6$ theory are known to the next order to that given above, \cite{12}. 
The various numbers which appear there are new nested sums which are not 
present even at four loops in $n$~$=$~$3$ and $4$. Instead in the latter the 
nested sums which arise are the usual Riemann $\zeta$-function at integer 
argument which lead to more complicated sums at much higher loops. For a 
variety of articles on this topic see, for instance, \cite{40,41,42,43,44}. 
Indeed we now know the coefficients of such new irrationals in the $O(N)$
$\phi^4$ $\beta$-function at six and seven loops, \cite{28,29,30}. As the 
appearance of the Riemann $\zeta$-function is as a consequence of the seeding 
by the hidden $\Gamma$-functions of integer argument, then by the same token we
would expect the appearance of new sums in the rational critical dimension 
theories seeded by combinations of $\Gamma$-functions with rational arguments. 

{\begin{figure}[hb]
\begin{center}
\includegraphics[width=8cm,height=2.0cm]{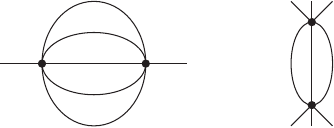}
\end{center}
\caption{Leading order topologies for $\phi^6$.} 
\label{lo6}
\end{figure}}

\sect{$O(N)$ symmetric theories.}

Having concentrated on the basic scalar theories with one field we now turn to
the case where the field has an $O(N)$ symmetry. Studies of $O(N)$ symmetric
theories in fractional dimensions have been carried out previously in 
\cite{16} for example. Here our motivation is to extend the renormalization 
group functions of theories considered in the previous section. One aim is to 
provide this information ahead of the application of conformal field theory 
ideas such as those developed in \cite{21} to $O(N)$ symmetric potentials. In 
\cite{21} conformal methods were used to compute critical exponents from which 
the renormalization group functions were constructed. A second reason is to 
highlight a possible connection between various theories with an $O(N)$ 
symmetry. To appreciate this it is perhaps best to recall a connection which 
has been widely studied. For instance, $O(N)$ symmetric $\phi^4$ theory can be 
described by the Lagrangian of bare quantities 
\begin{equation}
L_N^{\phi^4} ~=~ \frac{1}{2} \partial_\mu \phi^i_{\bare} 
\partial^\mu \phi^i_{\bare} ~+~
\frac{g_{\bare}}{2} \sigma_{\bare} \phi^i_{\bare} \phi^i_{\bare} ~-~ 
\frac{1}{2} \sigma_{\bare}^2 
\label{lag4n}
\end{equation}
where $\sigma_{\bare}$ is an auxiliary field here and is not to be confused 
with the exponent which was introduced briefly earlier and its associated 
renormalization constant here and elsewhere is
\begin{equation}
\sigma_{\bare} ~=~ \sqrt{Z_\sigma} \sigma ~.
\end{equation}
The elimination of $\sigma_{\bare}$ produces the canonical Lagrangian which is 
renormalizable in four dimensions. Recently it has been shown, \cite{33,34}, 
that the ultraviolet completion of $L_N^{\phi^4}$ is to the Lagrangian 
$L_N^{\phi^3}$ where 
\begin{equation}
L_N^{\phi^3} ~=~ \frac{1}{2} \partial_\mu \phi^i_{\bare} 
\partial^\mu \phi^i_{\bare} ~+~
\frac{1}{2} \partial^\mu \sigma_{\bare} \partial_\mu \sigma_{\bare} ~+~
\frac{g_{1\,\bare}}{2} \sigma_{\bare} \phi^i_{\bare} \phi^i_{\bare} ~+~ 
\frac{g_{2\,\bare}}{2} \sigma_{\bare}^3 ~.
\end{equation}
In $L_N^{\phi^3}$ $\sigma$ is not regarded now as an auxiliary field due to the 
usual kinetic term. Also its cubic self-interaction is required to ensure the
Lagrangian is renormalizable in six dimensions. The ultraviolet completion
relates to the fact that at the Wilson-Fisher fixed point in dimensions $d$
where $4$~$<$~$d$~$<$~$6$ both theories lie in the same universality class. 
This is not unrelated to the common interaction between the $\phi^i$ and 
$\sigma$ fields. Indeed one way to establish the universal connection is via
the $1/N$ expansion especially in the original formulation given in 
\cite{45,46,47}. In those articles the critical exponents of the universal 
theory were computed as functions of the spacetime to three terms in the $1/N$ 
expansion. Expanding the exponents in the neighbourhood of a critical dimension
of a theory such as $L_N^{\phi^3}$ and $L_N^{\phi^4}$ they are in full 
agreement with the critical exponents derived from the explicit renormalization
group functions of those theories. 

{\begin{figure}[ht]
\begin{center}
\includegraphics[width=10cm,height=10cm]{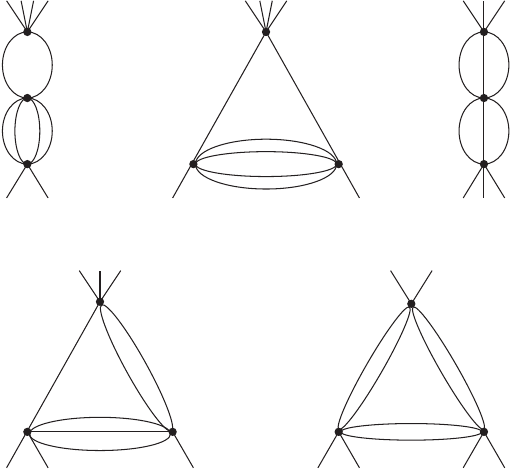}
\end{center}
\caption{Next to leading order topologies for $6$-point function in $\phi^6$ 
theory.} 
\label{vernlo6}
\end{figure}}

Having reviewed this well-established connection between $L_N^{\phi^3}$ and 
$L_N^{\phi^4}$ it is worth noting that the latter is in effect a theory with 
$n$~$=$~$4$ in contrast with the former which has $n$~$=$~$3$. Each is part of 
the Ising or Lee-Yang class of theories. It turns out that this connectivity 
extends to some of the theories we consider here. For instance, the next 
candidate theory to apply this concept to is that recently termed the 
Blume-Capel class. For instance, the parallel starting point to $L_N^{\phi^4}$ 
is $L_N^{\phi^8}$ where 
\begin{equation}
L_N^{\phi^8} ~=~ \frac{1}{2} \partial_\mu \phi^i_{\bare} 
\partial^\mu \phi^i_{\bare} ~+~
\frac{g^2_{\bare}}{40320} ( \phi^i_{\bare} \phi^i_{\bare} )^4 ~. 
\end{equation}
The interaction which is clearly in the $n$~$=$~$8$ class can be reformulated
with an auxiliary field $\sigma$ via
\begin{equation}
L_N^{\phi^8} ~=~ \frac{1}{2} \partial_\mu \phi^i_{\bare} 
\partial^\mu \phi^i_{\bare} ~+~
\frac{g_{\bare}}{24} \sigma_{\bare} ( \phi^i_{\bare} \phi^i_{\bare} )^2 ~-~ 
\frac{35}{2} \sigma_{\bare}^2 ~.
\label{lag8n}
\end{equation}
While the critical dimension is still $\frac{8}{3}$ the core interaction has an 
$n$~$=$~$5$ structure. Therefore to investigate whether there is a common
universality class underlying $L_N^{\phi^8}$ and an $O(N)$ symmetric 
$n$~$=$~$5$ theory parallel to the $L_N^{\phi^3}$ and $L_N^{\phi^4}$ one the 
first stage is to construct the corresponding renormalization group functions 
for $L_N^{\phi^8}$ and the potentially related one which we will term
$L_N^{\phi^5}$. Based on the commonality of the core interaction in the 
$\sigma$ field formulation of $L_N^{\phi^3}$ and $L_N^{\phi^4}$ we follow the
same prescription of building a renormalizable Lagrangian in $D_5$ based on the
$n$~$=$~$5$ $O(N)$ symmetric interaction of $L_N^{\phi^8}$. This produces
\begin{eqnarray}
L_N^{\phi^5} &=& \frac{1}{2} \partial_\mu \phi^i_{\bare} 
\partial^\mu \phi^i_{\bare} ~+~
\frac{1}{2} \partial_\mu \sigma_{\bare} \partial^\mu \sigma_{\bare} ~+~
\frac{g_{1\,\bare}}{24} \sigma_{\bare} ( \phi^i_{\bare} \phi^i_{\bare} )^2 ~+~
\frac{g_{2\,\bare}}{12} \sigma^3_{\bare} \phi^i_{\bare} \phi^i_{\bare} ~+~ 
\frac{g_{3\,\bare}}{120} \sigma_{\bare}^5 
\end{eqnarray}
which has quintic interactions and a propagating $\sigma$ field. Unlike 
$L_N^{\phi^3}$ there is an additional interaction between $\phi^i$ and $\sigma$
from renormalizability. While this potential connection will serve as a 
motivation for constructing the renormalization group functions we will also 
renormalize the following remaining $O(N)$ symmetric Lagrangians
\begin{eqnarray}
L_N^{\phi^6} &=& \frac{1}{2} \partial_\mu \phi^i_{\bare} 
\partial^\mu \phi^i_{\bare} ~+~
\frac{g_{\bare}}{720} ( \phi^i_{\bare} \phi^i_{\bare} )^3 \nonumber \\
L_N^{\phi^7} &=& \frac{1}{2} \partial_\mu \phi^i_{\bare} 
\partial^\mu \phi^i_{\bare} ~+~
\frac{1}{2} \partial_\mu \sigma_{\bare} \partial^\mu \sigma_{\bare} ~+~
\frac{g_{1\,\bare}}{720} \sigma_{\bare} ( \phi^i_{\bare} \phi^i_{\bare} )^3 ~+~
\frac{g_{2\,\bare}}{144} \sigma_{\bare}^3 ( \phi^i_{\bare} \phi^i_{\bare} )^2 
\nonumber \\
&& +~ \frac{g_{3\,\bare}}{240} \sigma_{\bare}^5 \phi^i_{\bare} 
\phi^i_{\bare} ~+~ \frac{g_{4\,\bare}}{5040} \sigma_{\bare}^7 \nonumber \\ 
L_N^{\phi^9} &=& \frac{1}{2} \partial_\mu \phi^i_{\bare}
\partial^\mu \phi^i_{\bare} ~+~
\frac{1}{2} \partial_\mu \sigma_{\bare} \partial^\mu \sigma_{\bare} ~+~
\frac{g_{1\,\bare}}{40320} \sigma_{\bare} 
( \phi^i_{\bare}\phi^i_{\bare})^4 ~+~
\frac{g_{2\,\bare}}{4320} \sigma_{\bare}^3 ( \phi^i_{\bare} \phi^i_{\bare} )^3 
\nonumber \\
&& +~ \frac{g_{3\,\bare}}{2880} \sigma_{\bare}^5 
( \phi^i_{\bare} \phi^i_{\bare} )^2 ~+~ 
\frac{g_{4\,\bare}}{10080} \sigma_{\bare}^7 \phi^i_{\bare} \phi^i_{\bare} ~+~ 
\frac{g_{5\,\bare}}{362880} \sigma_{\bare}^9 ~.
\end{eqnarray}
The Lagrangian $L_N^{\phi^6}$ was studied previously in \cite{12} and is 
similar in structure to $L_N^{\phi^8}$. By contrast for odd $n$ there are an
increasing number of interactions to ensure renormalizability. Consequently
there is a significantly larger number of Feynman graphs to be evaluated. We
have given an indication of this for the renormalization of the fields and 
coupling constants for Lagrangians with an odd value of $n$ in Table $2$.

{\begin{figure}[ht]
\begin{center}
\includegraphics[width=8cm,height=2.0cm]{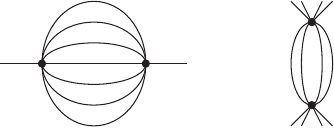}
\end{center}
\caption{Leading order topologies for $\phi^8$.} 
\label{verlo8}
\end{figure}}

{\begin{table}[ht]
\begin{center}
\begin{tabular}{|c||c|c|c|c|c|c|c|}
\hline
$n$ & $\phi^i$ & $\sigma$ & $g_1$ & $g_2$ & $g_3$ & $g_4$ & $g_5$ \\
\hline
$5$ & $2$ & $3$ & $135$ & $137$ & $155$ & --- & --- \\
$7$ & $3$ & $4$ & $2644$ & $2688$ & $2692$ & $2912$ & --- \\
$9$ & $4$ & $5$ & $42607$ & $43149$ & $43147$ & $43241$ & $42607$ \\
\hline
\end{tabular}
\end{center}
\begin{center}
{Table 2. Number of graphs computed for the renormalization of the field and
each coupling constant for $O(N)$ symmetric theories for odd $n$.}
\end{center}
\end{table}}

In light of this we note that the Lagrangian for higher order interaction 
equivalences is straightforward to write down. The Lagrangian $L_N^{\phi^{4r}}$ 
should have connection with $L_N^{\phi^{2r+1}}$ for integer $r$ where
\begin{equation}
L_N^{\phi^{4r}} ~=~ \frac{1}{2} \partial_\mu \phi^i_{\bare}  
\partial^\mu \phi^i_{\bare}  ~+~
\frac{g_{\bare} }{(4r)!} ( \phi^i_{\bare}  \phi^i_{\bare}  )^{2r} 
\end{equation}
and $L_N^{\phi^{2r+1}}$ has the core interaction $\sigma (\phi^i \phi^i)^r$
leading to
\begin{equation}
L_N^{\phi^{2r+1}} ~=~ \frac{1}{2} \partial_\mu \phi^i_{\bare}  
\partial^\mu \phi^i_{\bare} ~+~
\frac{1}{2} \partial_\mu \sigma_{\bare} \partial^\mu \sigma_{\bare} ~+~
\sum_{p=0}^r g_{p+1\,\bare} \sigma_{\bare}^{2p+1} ( \phi^i_{\bare} 
\phi^i_{\bare} )^{r-p} ~.
\end{equation}
In making this connection with various odd and even potentials it is 
straightforward to write down the Lagrangian of the limiting critical 
dimension. To see the connection between $L_N^{\phi^{2r+1}}$ and 
$L_N^{\phi^{4r}}$ via the same $n$-point vertex we introduce the auxiliary 
field $\sigma$ as in (\ref{lag4n}) and (\ref{lag8n}). In the former there is a 
theory with a lower critical dimension which is in the same universality class 
as $L_N^{\phi^4}$ and is the nonlinear $\sigma$ model (nlsm) which has a 
Lagrangian and can be written as
\begin{equation}
L_N^{\mbox{\footnotesize{nlsm}}} ~=~ 
\frac{1}{2} \partial_\mu \phi^i_{\bare} \partial^\mu \phi^i_{\bare} ~+~
\frac{g_{\bare}}{2} \sigma_{\bare} \phi^i_{\bare} \phi^i_{\bare} ~-~ 
\frac{1}{2} \sigma_{\bare} ~.
\label{lag2n}
\end{equation}
Here $\sigma$ is regarded as a Lagrange multiplier field and its role is to 
constrain the fields $\phi^i$ to lie on a sphere. Also the critical dimension
of (\ref{lag2n}) is $2$ and in this dimension $\sigma$ has canonical dimension 
$2$ whereas the $\phi^i$ field is dimensionless. However with parallel 
Lagrangians based on higher order potentials also available one can write down
similar Lagrangians which are linear in $\sigma$ and which have critical
dimension $2$. This can be generalized to the Lagrangian
\begin{equation}
L_N^{\phi^\infty} ~=~ 
\frac{1}{2} \partial_\mu \phi^i_{\bare} \partial^\mu \phi^i_{\bare} ~+~
\frac{1}{2} \sum_{n=0}^\infty g_{n\,\bare} \sigma_{\bare} 
( \phi^i_{\bare} \phi^i_{\bare} )^n 
\label{laginftyn}
\end{equation}
where all possible interactions are available and has $D_\infty$~$=$~$2$. One 
can regard (\ref{laginftyn}) as the theory corresponding to (\ref{critd}) in 
the $n$~$\to$~$\infty$ limit. Equally it could be viewed as the base two
dimensional Lagrangian from which each of the $D_n$~$>$~$2$ theories we have 
considered here, as well as others, are related to through their corresponding 
Wilson-Fisher fixed point. In some sense it is the two dimensional universal
Lagrangian of all the univerality classes of scalar $O(N)$ theories. The
connection of each interaction relative to its critical dimension and two
dimensions is apparent if one considers the structure of the leading order
graph contributing to the $2$-point function renormalization such as those
illustrated in Figures \ref{2ptlo57}, \ref{lo6} and \ref{verlo8} for 
example\footnote{The author is indebted to Prof D. Kreimer for the reminder 
about the structure of this set of graphs.}. If we denote such a graph with $L$
loops by $\Gamma_{(2),L}$ then its evaluation is
\begin{equation}
\Gamma_{(2),L} ~=~ \frac{\Gamma^{L+1}\left(\frac{d}{2}-1\right)
\Gamma\left(L+1-\frac{d}{2}L\right)} 
{\Gamma\left(\frac{d}{2}(L+1)-(L+1)\right)} 
\end{equation}
which corresponds to a $\phi^{L+2}$ potential. This function diverges when the
argument of either numerator $\Gamma$-functions is zero or negative. Clearly
this will occur for all $L$~$\geq$~$1$ when $d$~$=$~$2$ due to
$\Gamma\left(\frac{d}{2}-1\right)$. For lower values of $d$ this is not
meaningful. By contrast the other numerator $\Gamma$-function diverges for
certain dimensions but this depends on $L$ and hence the specific potential.
More crucially for (\ref{laginftyn}) the critical dimensions $D_n$ emerge for
each potential when this $\Gamma$-function has argument $(-1)$. Therefore at
the initial renormalization stage the Lagrangian (\ref{laginftyn}) reflects the
dimensional connection. Of course there are other singularities in this
specific $\Gamma$-function when the argument is any other negative integer or
zero but these do not correspond to the critical dimensions of any of the
Lagrangians we consider here. Instead several may correspond to their higher
dimensional ultraviolet completions.

{\begin{figure}[ht]
\begin{center}
\includegraphics[width=16cm,height=10cm]{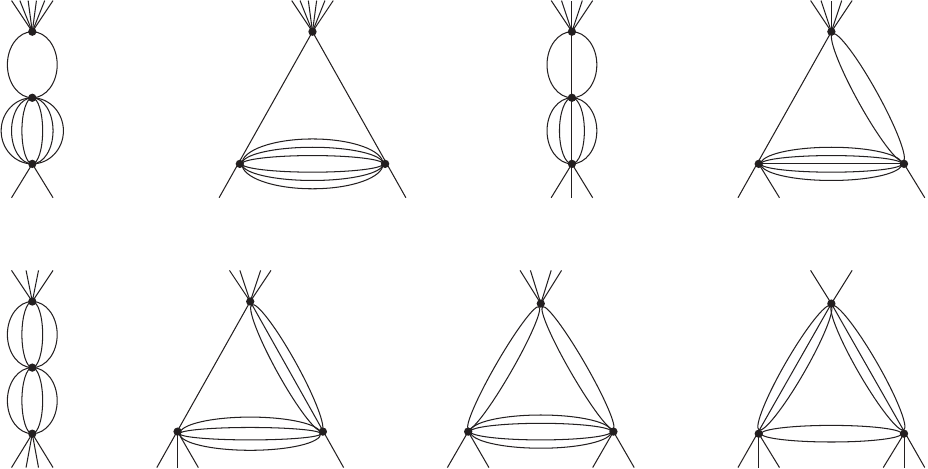}
\end{center}
\caption{Next to leading order topologies for $8$-point function in $\phi^8$ 
theory.} 
\label{vernlo8}
\end{figure}}

We now proceed to the task of recording the results. First, for the potential
connection between $L_N^{\phi^5}$ and $L_N^{\phi^8}$ we have  
\begin{eqnarray}
\gamma_\phi^{\phi^5}(g_i) &=& -~ \left[ N g_1^2 + 2 g_1^2 + 3 g_2^2 \right]
\frac{\Gamma^3 \left( \frac{2}{3} \right)}{20} ~+~ O(g_i^3) \nonumber \\
\gamma_\sigma^{\phi^5}(g_i) &=& -~ \left[ N^2 g_1^2 + 2 N g_1^2 + 18 N g_2^2
+ 3 g_3^2 \right] \frac{\Gamma^3 \left( \frac{2}{3} \right)}{80}~+~ O(g_i^3)
\nonumber \\
\beta_1^{\phi^5}(g_i) &=& -~ 3 \epsilon g_1 \nonumber \\
&& +~ \left[ 
59 N^2 g_1^3 + 2182 N g_1^3 + 9648 g_1^3 + 1440 N g_1^2 g_2 + 15840 g_1^2 g_2 
+ 1062 N g_1 g_2^2 
\right. \nonumber \\
&& \left. ~~~~ 
+ 12912 g_1 g_2^2 - 480 g_1 g_2 g_3 - 3 g_1 g_3^2 + 10800 g_2^3 
+ 1620 g_2^2 g_3 \right]
\frac{\Gamma^3 \left( \frac{2}{3} \right)}{80} ~+~ O(g_i^5) \nonumber \\
\beta_2^{\phi^5}(g_i) &=& -~ 3 \epsilon g_2 \nonumber \\
&& +~
\left[ 80 N^2 g_1^3 + 1040 N g_1^3 + 1760 g_1^3 + 177 N^2 g_1^2 g_2 
+ 2506 N g_1^2 g_2 + 4304 g_1^2 g_2 
\right. \nonumber \\
&& \left. ~~~~
- 40 N g_1^2 g_3 - 80 g_1^2 g_3 + 5400 N g_1 g_2^2 + 10800 g_1 g_2^2 
+ 540 N g_1 g_2 g_3 
\right. \nonumber \\
&& \left. ~~~~
+ 1080 g_1 g_2 g_3 + 486 N g_2^3 + 17136 g_2^3 + 8640 g_2^2 g_3 
+ 1251 g_2 g_3^2 \right]
\frac{\Gamma^3 \left( \frac{2}{3} \right)}{80} ~+~ O(g_i^5) \nonumber \\
\beta_3^{\phi^5}(g_i) &=& -~ 3 \epsilon g_3 \nonumber \\
&& +~
\left[ 
-~ 80 N^2 g_1^2 g_2 - 160 N g_1^2 g_2 - N^2 g_1^2 g_3 - 2 N g_1^2 g_3 
+ 540 N^2 g_1 g_2^2 + 1080 N g_1 g_2^2 
\right. \nonumber \\
&& \left. ~~~~
+ 5760 N g_2^3 + 2502 N g_2^2 g_3 + 1377 g_3^3 \right]
\frac{\Gamma^3 \left( \frac{2}{3} \right)}{16} ~+~ O(g_i^5) 
\end{eqnarray}
and 
\begin{eqnarray}
\gamma_\phi^{\phi^8}(g) &=& 
\frac{2\pi^6}{99225 \Gamma^6 \left( \frac{2}{3} \right) }
[N+2] [N+4] [N+6] g^2 \nonumber \\
&& -~ \frac{\pi^9 \sqrt{3}}{20837250 \Gamma^9 \left( \frac{2}{3} \right)}
[N+2] [N+4] [N+6] [ 3 N^2 + 150 N + 1072 ] g^3 ~+~ O(g^4) \nonumber \\
\gamma_{\cal O}^{\phi^8}(g) &=& 
\frac{26 \pi^6}{99225 \Gamma^6 \left( \frac{2}{3} \right) }
[N+2] [N+4] [N+6] g^2 ~+~ O(g^3) \nonumber \\ 
\beta^{\phi^8}(g) &=& -~ 6 \epsilon g ~+~ 
\frac{2\pi^3 \sqrt{3}}{\Gamma^3 \left( \frac{2}{3} \right)}
\left[
\frac{1072}{315} + \frac{10}{21} N + \frac{1}{105} N^2 \right] g^2 \nonumber \\
&& + \left[ 
-~ \frac{\pi^6}{\Gamma^6 \left( \frac{2}{3} \right)}
\left[
\frac{164480}{441}
+ \frac{74464}{735} N
+ \frac{4576}{525} N^2
+ \frac{184}{735} N^3
+ \frac{8}{3675} N^4 \right] 
\right. \nonumber \\
&& \left. ~~~~
- \frac{\pi^6\ln{3}}{\Gamma^6 \left( \frac{2}{3} \right)}
\left[
\frac{26496}{245}
+ \frac{15392}{525} N
+ \frac{9056}{3675} N^2
+ \frac{256}{3675} N^3
+ \frac{4}{3675} N^4 \right] 
\right. \nonumber \\
&& \left. ~~~~
- \frac{\pi^9\sqrt{3}}{\Gamma^9 \left( \frac{2}{3} \right)}
\left[
\frac{11558912}{297675}
+ \frac{1712128}{178605} N
+ \frac{217088}{297675} N^2
+ \frac{1024}{59535} N^3 \right] 
\right. \nonumber \\
&& \left. ~~~~
+ \frac{\pi^7\sqrt{3}}{\Gamma^6 \left( \frac{2}{3} \right)}
\left[
\frac{8832}{245}
+ \frac{15392}{1575} N
+ \frac{9056}{11025} N^2
+ \frac{256}{11025} N^3
+ \frac{4}{11025} N^4 \right] 
\right] g^3 \nonumber \\
&& +~ O(g^4) 
\label{rged8N}
\end{eqnarray}
where the order symbol for the multi-coupling theories denotes all possible
combinations of the couplings. We have also recorded the mass operator 
dimension in passing. While the $\beta$-function in (\ref{rged8N}) is not
asymptotically free there is a Banks-Zaks fixed point, \cite{48} for all $N$. 
Given the proximity of the critical dimension of $\phi^8$ theory to two 
dimensions it ought to be possible to use the $\epsilon$-expansion to estimate
the wave function critical exponent in that lower spacetime. For instance, when
$N$~$=$~$1$ it is known, \cite{13}, that $\phi^8$ theory corresponds to the 
unitary, conformal, minimal model with $c$~$=$~$\frac{4}{5}$. However, it was 
shown in \cite{13} that a sizeable number of terms of the $\epsilon$-expansion 
would be required to have approximate agreement.

For the remaining two theories we are concentrating on we have 
\begin{eqnarray}
\gamma_\phi^{\phi^7}(g_i) &=& -~
\left[ \frac{5}{336} g_3^2
+ \frac{25}{756} g_2^2
+ \frac{1}{126} g_1^2
+ \frac{25}{1512} N g_2^2
+ \frac{1}{168} N g_1^2
+ \frac{1}{1008} N^2 g_1^2
\right]
\Gamma^5 \left( \frac{2}{5} \right) \nonumber \\
&& +~ O(g_i^3) \nonumber \\
\gamma_\sigma^{\phi^7}(g_i) &=& -~
\left[
\frac{5}{2016} g_4^2
+ \frac{25}{672} N g_3^2
+ \frac{25}{1008} N g_2^2
+ \frac{1}{756} N g_1^2
+ \frac{25}{2016} N^2 g_2^2
+ \frac{1}{1008} N^2 g_1^2
\right. \nonumber \\
&& \left. ~~~~
+ \frac{1}{6048} N^3 g_1^2
\right]
\Gamma^5 \left( \frac{2}{5} \right) ~+~ O(g_i^3) \nonumber \\
\beta_1^{\phi^7}(g_i) &=& -~ 5 \epsilon g_1 \nonumber \\
&& + \left[
- \frac{625}{48} g_2 g_3 g_4
+ \frac{9875}{24} g_2 g_3^2
+ \frac{625}{18} g_2^2 g_4
+ \frac{8125}{12} g_2^2 g_3
+ \frac{18625}{18} g_2^3
- \frac{5}{2016} g_1 g_4^2
\right. \nonumber \\
&& \left. ~~~~
- \frac{5}{12} g_1 g_3 g_4
- \frac{460}{7} g_1 g_3^2
+ \frac{1225}{9} g_1 g_2 g_3
+ \frac{246275}{189} g_1 g_2^2
+ 515 g_1^2 g_2
+ \frac{18932}{63} g_1^3
\right. \nonumber \\
&& \left. ~~~~
- \frac{625}{24} N g_2^2 g_3
+ \frac{3625}{36} N g_2^3
- \frac{1775}{672} N g_1 g_3^2
+ \frac{1225}{18} N g_1 g_2 g_3
+ \frac{471775}{3024} N g_1 g_2^2
\right. \nonumber \\
&& \left. ~~~~
+ \frac{395}{4} N g_1^2 g_2
+ \frac{8165}{108} N g_1^3
- \frac{5275}{2016} N^2 g_1 g_2^2
+ \frac{65}{8} N^2 g_1^2 g_2
+ \frac{647}{144} N^2 g_1^3
\right. \nonumber \\
&& \left. ~~~~
- \frac{211}{6048} N^3 g_1^3
\right]
\Gamma^5 \left( \frac{2}{5} \right)
\nonumber \\
&&
+
\left[
\frac{375}{8} g_3^3
+ \frac{125} g_2 g_3^2
+ \frac{1750}{3} g_2^2 g_3
+ 875 g_2^3
+ \frac{350}{3} g_1 g_2 g_3
+ \frac{12875}{18} g_1 g_2^2
+ 470 g_1^2 g_2
\right. \nonumber \\
&& \left. ~~~~
+ \frac{632}{3} g_1^3
+ \frac{625}{24} N g_2^2 g_3
+ \frac{125}{2} N g_2^3
+ \frac{25}{3} N g_1 g_2 g_3
+ \frac{1325}{12} N g_1 g_2^2
+ 90 N g_1^2 g_2
\right. \nonumber \\
&& \left. ~~~~
+ \frac{311}{6} N g_1^3
+ \frac{175}{72} N^2 g_1 g_2^2
+ \frac{5}{2} N^2 g_1^2 g_2
+ \frac{37}{12} N^2 g_1^3
+ \frac{1}{24} N^3 g_1^3
\right]
\frac{\Gamma^6 \left( \frac{2}{5} \right)\Gamma^2 \left( \frac{3}{5} \right)}
{\Gamma^2 \left( \frac{4}{5} \right)} \nonumber \\
&& +~ O(g_i^5) \nonumber \\
\beta_2^{\phi^7}(g_i) &=& -~ 5 \epsilon g_2 \nonumber \\
&& + \left[
\frac{375}{8} g_3^2 g_4
+ \frac{2725}{12} g_3^3
- \frac{1895}{672} g_2 g_4^2
+ \frac{245}{6} g_2 g_3 g_4
+ \frac{18245}{28} g_2 g_3^2
+ \frac{3475}{3} g_2^2 g_3
\right. \nonumber \\
&& \left. ~~~~
+ \frac{492725}{567} g_2^3
- \frac{25}{12} g_1 g_3 g_4
+ \frac{395}{6} g_1 g_3^2
+ \frac{100}{9} g_1 g_2 g_4
+ \frac{650}{3} g_1 g_2 g_3
+ \frac{1490}{3} g_1 g_2^2
\right. \nonumber \\
&& \left. ~~~~
+ \frac{98}{9} g_1^2 g_3
+ \frac{39404}{189} g_1^2 g_2
+ \frac{412}{15} g_1^3
+ \frac{47525}{672} N g_2 g_3^2
+ \frac{425}{3} N g_2^2 g_3
+ \frac{1869475}{9072} N g_2^3
\right. \nonumber \\
&& \left. ~~~~
- \frac{25}{48} N g_1 g_3 g_4
+ \frac{395}{24} N g_1 g_3^2
+ \frac{25}{9} N g_1 g_2 g_4
+ \frac{275}{6} N g_1 g_2 g_3
+ \frac{345}{2} N g_1 g_2^2
\right. \nonumber \\
&& \left. ~~~~
+ \frac{49}{6} N g_1^2 g_3
+ \frac{925}{12} N g_1^2 g_2
+ \frac{182}{15} N g_1^3
+ \frac{10825}{672} N^2 g_2^3
- \frac{25}{12} N^2 g_1 g_2 g_3
\right. \nonumber \\
&& \left. ~~~~
+ \frac{145}{12} N^2 g_1 g_2^2
+ \frac{49}{36} N^2 g_1^2 g_3
+ \frac{2515}{432} N^2 g_1^2 g_2
+ \frac{7}{4} N^2 g_1^3
- \frac{211}{2016} N^3 g_1^2 g_2
\right. \nonumber \\
&& \left. ~~~~
+ \frac{13}{120} N^3 g_1^3
\right]
\Gamma^5 \left( \frac{2}{5} \right) \nonumber \\
&&
+
\left[
\frac{175}{8} g_3^2 g_4
+ \frac{375}{2} g_3^3
+ 25 g_2 g_3 g_4
+ \frac{3025}{6} g_2 g_3^2
+ 25 g_2^2 g_4
+ 775 g_2^2 g_3
+ \frac{32975}{54} g_2^3
\right. \nonumber \\
&& \left. ~~~~
+ 20 g_1 g_3^2
+ \frac{560}{3} g_1 g_2 g_3
+ 420 g_1 g_2^2
+ \frac{28}{3} g_1^2 g_3
+ \frac{1030}{9} g_1^2 g_2
+ \frac{376}{15} g_1^3
+ \frac{325}{12} N g_2 g_3^2
\right. \nonumber \\
&& \left. ~~~~
+ \frac{25}{8} N g_2^2 g_4
+ 125 N g_2^2 g_3
+ \frac{5275}{36} N g_2^3
+ 5 N g_1 g_3^2
+ 55 N g_1 g_2 g_3
+ 135 N g_1 g_2^2
\right. \nonumber \\
&& \left. ~~~~
+ 3 N g_1^2 g_3
+ \frac{833}{18} N g_1^2 g_2
+ \frac{166}{15} N g_1^3
+ \frac{475}{216} N^2 g_2^3
+ \frac{25}{12} N^2 g_1 g_2 g_3
+ \frac{15}{2} N^2 g_1 g_2^2
\right. \nonumber \\
&& \left. ~~~~
+ \frac{1}{6} N^2 g_1^2 g_3
+ \frac{173}{36} N^2 g_1^2 g_2
+ \frac{4}{3} N^2 g_1^3
+ \frac{7}{72} N^3 g_1^2 g_2
+ \frac{1}{30} N^3 g_1^3
\right]
\frac{\Gamma^6 \left( \frac{2}{5} \right)\Gamma^2 \left( \frac{3}{5} \right)}
{\Gamma^2 \left( \frac{4}{5} \right)} \nonumber \\
&& +~ O(g_i^5) \nonumber \\
\beta_3^{\phi^7}(g_i) &=& -~ 5 \epsilon g_3 \nonumber \\
&& + \left[
\frac{25775}{672} g_3 g_4^2
+ \frac{4975}{24} g_3^2 g_4
+ \frac{21775}{28} g_3^3
+ \frac{625}{6} g_2 g_3 g_4
+ \frac{13625}{18} g_2 g_3^2
+ \frac{1225}{54} g_2^2 g_4
\right. \nonumber \\
&& \left. ~~~~
+ \frac{91225}{126} g_2^2 g_3
+ \frac{34750}{81} g_2^3
- \frac{125}{54} g_1 g_2 g_4
+ \frac{3950}{27} g_1 g_2 g_3
+ \frac{3250}{27} g_1 g_2^2
- \frac{1}{27} g_1^2 g_4
\right. \nonumber \\
&& \left. ~~~~
- \frac{736}{63} g_1^2 g_3
+ \frac{980}{81} g_1^2 g_2
+ \frac{81875}{2016} N g_3^3
+ \frac{625}{12} N g_2 g_3 g_4
+ \frac{13625}{36} N g_2 g_3^2
\right. \nonumber \\
&& \left. ~~~~
+ \frac{1225}{108} N g_2^2 g_4
+ \frac{1332325}{3024} N g_2^2 g_3
+ \frac{21625}{81} N g_2^3
- \frac{125}{72} N g_1 g_2 g_4
+ \frac{1975}{18} N g_1 g_2 g_3
\right. \nonumber \\
&& \left. ~~~~
+ \frac{4625}{54} N g_1 g_2^2
- \frac{1}{36} N g_1^2 g_4
- \frac{997}{108} N g_1^2 g_3
+ \frac{1225}{81} N g_1^2 g_2
+ \frac{237625}{6048} N^2 g_2^2 g_3
\right. \nonumber \\
&& \left. ~~~~
+ \frac{2125}{81} N^2 g_2^3
- \frac{125}{432} N^2 g_1 g_2 g_4
+ \frac{1975}{108} N^2 g_1 g_2 g_3
+ \frac{625}{54} N^2 g_1 g_2^2
- \frac{1}{216} N^2 g_1^2 g_4
\right. \nonumber \\
&& \left. ~~~~
- \frac{29}{16} N^2 g_1^2 g_3
+ \frac{490}{81} N^2 g_1^2 g_2
- \frac{125}{216} N^3 g_1 g_2^2
- \frac{355}{6048} N^3 g_1^2 g_3
\right. \nonumber \\
&& \left. ~~~~
+ \frac{245}{324} N^3 g_1^2 g_2
\right]
\Gamma^5 \left( \frac{2}{5} \right) \nonumber \\
&&
+
\left[
\frac{625}{24} g_3 g_4^2
+ \frac{375}{2} g_3^2 g_4
+ \frac{1375}{3} g_3^3
+ \frac{875}{18} g_2 g_3 g_4
+ 625 g_2 g_3^2
+ \frac{125}{9} g_2^2 g_4
+ \frac{15125}{27} g_2^2 g_3
\right. \nonumber \\
&& \left. ~~~~
+ \frac{7750}{27} g_2^3
+ 25 g_1 g_3^2
+ \frac{400}{9} g_1 g_2 g_3
+ \frac{2800}{27} g_1 g_2^2
+ \frac{280}{27} g_1^2 g_2
+ \frac{1375}{24} N g_3^3
\right. \nonumber \\
&& \left. ~~~~
+ \frac{875}{36} N g_2 g_3 g_4
+ \frac{625}{2} N g_2 g_3^2
+ \frac{125}{18} N g_2^2 g_4
+ \frac{8375}{27} N g_2^2 g_3
+ \frac{5125}{27} N g_2^3
\right. \nonumber \\
&& \left. ~~~~
+ \frac{75}{4} N g_1 g_3^2
+ \frac{100}{3} N g_1 g_2 g_3
+ \frac{2225}{27} N g_1 g_2^2
+ \frac{230}{27} N g_1^2 g_2
+ \frac{1625}{108} N^2 g_2^2 g_3
\right. \nonumber \\
&& \left. ~~~~
+ \frac{625}{27} N^2 g_2^3
+ \frac{25}{8} N^2 g_1 g_3^2
+ \frac{50}{9} N^2 g_1 g_2 g_3
+ \frac{1775}{108} N^2 g_1 g_2^2
+ \frac{50}{27} N^2 g_1^2 g_2
\right. \nonumber \\
&& \left. ~~~~
+ \frac{125}{216} N^3 g_1 g_2^2
+ \frac{5}{54} N^3 g_1^2 g_2
\right]
\frac{\Gamma^6 \left( \frac{2}{5} \right)\Gamma^2 \left( \frac{3}{5} \right)}
{\Gamma^2 \left( \frac{4}{5} \right)} ~+~ O(g_i^5) \nonumber \\
\beta_4^{\phi^7}(g_i) &=& -~ 5 \epsilon g_4 \nonumber \\
&& + \left[
\frac{43325}{288} g_4^3
+ \frac{25775}{32} N g_3^2 g_4
+ \frac{34825}{24} N g_3^3
+ \frac{4375}{4} N g_2 g_3^2
- \frac{9475}{144} N g_2^2 g_4
\right. \nonumber \\
&& \left. ~~~~
+ \frac{8575}{18} N g_2^2 g_3
- \frac{875}{18} N g_1 g_2 g_3
+ \frac{3500}{27} N g_1 g_2^2
- \frac{1}{108} N g_1^2 g_4
- \frac{7}{9} N g_1^2 g_3
\right. \nonumber \\
&& \left. ~~~~
+ \frac{4375}{8} N^2 g_2 g_3^2
- \frac{9475}{288} N^2 g_2^2 g_4
+ \frac{8575}{36} N^2 g_2^2 g_3
- \frac{875}{24} N^2 g_1 g_2 g_3
+ \frac{875}{9} N^2 g_1 g_2^2
\right. \nonumber \\
&& \left. ~~~~
- \frac{1}{144} N^2 g_1^2 g_4
- \frac{7}{12} N^2 g_1^2 g_3
- \frac{875}{144} N^3 g_1 g_2 g_3
+ \frac{875}{54} N^3 g_1 g_2^2
- \frac{1}{864} N^3 g_1^2 g_4
\right. \nonumber \\
&& \left. ~~~~
- \frac{7}{72} N^3 g_1^2 g_3
\right]
\Gamma^5 \left( \frac{2}{5} \right) \nonumber \\
&& +
\left[
\frac{875}{8} g_4^3
+ \frac{4375}{8} N g_3^2 g_4
+ \frac{2625}{2} N g_3^3
+ \frac{6125}{12} N g_2 g_3^2
+ \frac{875}{3} N g_2^2 g_3
+ \frac{1750}{9} N g_2^3
\right. \nonumber \\
&& \left. ~~~~
+ \frac{6125}{24} N^2 g_2 g_3^2
+ \frac{875}{6} N^2 g_2^2 g_3
+ \frac{4375}{36} N^2 g_2^3
+ \frac{875}{72} N^3 g_2^3
\right]
\frac{\Gamma^6 \left( \frac{2}{5} \right)\Gamma^2 \left( \frac{3}{5} \right)}
{\Gamma^2 \left( \frac{4}{5} \right)} \nonumber \\
&& +~ O(g_i^5)
\end{eqnarray}
for $L_N^{\phi^7}$. For $L_N^{\phi^9}$ the anomalous dimensions are 
\begin{eqnarray}
\gamma_\phi^{\phi^9}(g_i) &=& -~ \left[ N^3 g_1^2 + 12 N^2 g_1^2 + 44 N g_1^2 
+ 48 g_1^2 + 49 N^2 g_2^2  + 294 N g_2^2 + 392 g_2^2
\right. \nonumber \\
&& \left. ~~~~ 
+ 245 N g_3^2 + 490 g_3^2 + 105 g_4^2 \right] 
\frac{\Gamma^7 \left( \frac{2}{7} \right)}{194400} ~+~ O(g_i^3) \nonumber \\
\gamma_\sigma^{\phi^9}(g_i) &=& -~ \left[ N^4 g_1^2 + 12 N^3 g_1^2 
+ 44 N^2 g_1^2 + 48 N g_1^2 + 196 N^3 g_2^2
+ 1176 N^2 g_2^2 + 1568 N g_2^2 
\right. \nonumber \\
&& \left. ~~~~ 
+ 2450 N^2 g_3^2 + 4900 N g_3^2
+ 2940 N g_4^2 + 105 g_5^2 \right] 
\frac{\Gamma^7 \left( \frac{2}{7} \right)}{1555200} ~+~ O(g_i^3) ~. 
\end{eqnarray}
As there are five couplings for $L_N^{\phi^9}$ we record only one
$\beta$-function explicitly which will suffice for discussion purposes. The
remaining $\beta$-functions together with the renormalization group functions
for this and all the other Lagrangians are given in the attached data file. We
have 
\begin{eqnarray}
\beta_1^{\phi^9}(g_i) &=& -~ 7 \epsilon g_1 \nonumber \\
&& +~
\left[
\frac{12005}{576} g_3^2 g_5
+ \frac{12005}{27} g_3^2 g_4
- \frac{2401}{5400} g_2 g_4 g_5
- \frac{81977}{900} g_2 g_4^2
- \frac{2401}{270} g_2 g_3 g_5
\right. \nonumber \\
&& \left. ~~~~
+ \frac{602651}{1350} g_2 g_3 g_4
+ \frac{2020613}{810} g_2 g_3^2
+ \frac{33614}{135} g_2^2 g_4
+ \frac{1745527}{675} g_2^2 g_3
+ \frac{6269354}{3375} g_2^3
\right. \nonumber \\
&& \left. ~~~~
- \frac{7}{11520} g_1 g_5^2
- \frac{7}{450} g_1 g_4 g_5
- \frac{1568}{675} g_1 g_4^2
- \frac{37877}{675} g_1 g_3 g_4
+ \frac{900473}{8100} g_1 g_3^2
\right. \nonumber \\
&& \left. ~~~~
+ \frac{1138172}{2025} g_1 g_2 g_3
+ \frac{7829612}{3375} g_1 g_2^2
+ \frac{174412}{375} g_1^2 g_2
+ \frac{933413}{3375} g_1^3
\right. \nonumber \\
&& \left. ~~~~
+ \frac{103243}{2160} N g_2 g_3 g_4
+ \frac{263081}{1620} N g_2 g_3^2
- \frac{2401}{540} N g_2^2 g_4
+ \frac{247303}{1080} N g_2^2 g_3
\right. \nonumber \\
&& \left. ~~~~
+ \frac{695261}{2250} N g_2^3
- \frac{3479}{43200} N g_1 g_4^2
- \frac{11123}{1350} N g_1 g_3 g_4
+ \frac{7284977}{129600} N g_1 g_3^2
\right. \nonumber \\
&& \left. ~~~~
+ \frac{15337}{135} N g_1 g_2 g_3
+ \frac{33848857}{81000} N g_1 g_2^2
+ \frac{2751}{25} N g_1^2 g_2
+ \frac{12077399}{162000} N g_1^3
\right. \nonumber \\
&& \left. ~~~~
+ \frac{103243}{10800} N^2 g_2^2 g_3
+ \frac{195853}{13500} N^2 g_2^3
+ \frac{55517}{51840} N^2 g_1 g_3^2
- \frac{1519}{4050} N^2 g_1 g_2 g_3
\right. \nonumber \\
&& \left. ~~~~
+ \frac{2276981}{108000} N^2 g_1 g_2^2
+ \frac{6601}{750} N^2 g_1^2 g_2
+ \frac{170927}{25920} N^2 g_1^3
+ \frac{173509}{648000} N^3 g_1 g_2^2
\right. \nonumber \\
&& \left. ~~~~
+ \frac{7}{50} N^3 g_1^2 g_2
+ \frac{132319}{648000} N^3 g_1^3
+ \frac{3541}{2592000} N^4 g_1^3
\right]
\Gamma^7 \left( \frac{2}{7} \right) 
\nonumber \\
&& +~
\left[ 
-~ \frac{2401}{48} g_3 g_4^2
- \frac{12005}{27} g_3^3
- \frac{2401}{30} g_2 g_4^2
- \frac{88837}{180} g_2 g_3 g_4
- \frac{45619}{27} g_2 g_3^2
\right. \nonumber \\
&& \left. ~~~~
- \frac{463393}{270} g_2^2 g_3
- \frac{451388}{225} g_2^3
- \frac{686}{9} g_1 g_3 g_4
- \frac{74431}{180} g_1 g_3^2
- \frac{93296}{135} g_1 g_2 g_3
\right. \nonumber \\
&& \left. ~~~~
- \frac{1026256}{675} g_1 g_2^2
- \frac{11368}{25} g_1^2 g_2
- \frac{18802}{75} g_1^3
- \frac{12005}{216} N g_3^3
- \frac{2401}{360} N g_2 g_3 g_4
\right. \nonumber \\
&& \left. ~~~~
- \frac{12005}{108} N g_2 g_3^2
- \frac{117649}{360} N g_2^2 g_3
- \frac{9604}{25} N g_2^3
- \frac{343}{90} N g_1 g_3 g_4
- \frac{7889}{216} N g_1 g_3^2
\right. \nonumber \\
&& \left. ~~~~
- \frac{4802}{45} N g_1 g_2 g_3
- \frac{45962}{135} N g_1 g_2^2
- \frac{10094}{75} N g_1^2 g_2
- \frac{32669}{450} N g_1^3
- \frac{16807}{2160} N^2 g_2^2 g_3
\right. \nonumber \\
&& \left. ~~~~
- \frac{2401}{225} N^2 g_2^3
- \frac{343}{2160} N^2 g_1 g_3^2
- \frac{343}{135} N^2 g_1 g_2 g_3
- \frac{10633}{540} N^2 g_1 g_2^2
- \frac{539}{50} N^2 g_1^2 g_2
\right. \nonumber \\
&& \left. ~~~~
- \frac{2989}{450} N^2 g_1^3
- \frac{343}{1350} N^3 g_1 g_2^2
- \frac{49}{300} N^3 g_1^2 g_2
- \frac{91}{450} N^3 g_1^3
\right. \nonumber \\
&& \left. ~~~~
- \frac{7}{3600} N^4 g_1^3
\right]
\frac{\Gamma^8 \left( \frac{2}{7} \right)
\Gamma^2 \left( \frac{5}{7} \right) \Gamma \left( \frac{6}{7} \right)}
{\Gamma \left( \frac{3}{7} \right) \Gamma^2 \left( \frac{4}{7} \right)}
\nonumber \\
&& +~
\left[ 
\frac{1715}{24} g_3 g_4^2
+ \frac{3430}{9} g_3^2 g_4
+ \frac{13720}{9} g_3^3
+ \frac{9947}{18} g_2 g_3 g_4
+ \frac{35672}{9} g_2 g_3^2
+ \frac{2744}{15} g_2^2 g_4
\right. \nonumber \\
&& \left. ~~~~
+ \frac{249361}{45} g_2^2 g_3
+ \frac{373184}{75} g_2^3
+ \frac{4949}{18} g_1 g_3^2
+ \frac{59584}{45} g_1 g_2 g_3
+ \frac{163072}{45} g_1 g_2^2
\right. \nonumber \\
&& \left. ~~~~
+ \frac{34496}{25} g_1^2 g_2
+ \frac{4852}{9} g_1^3
+ \frac{1715}{36} N g_3^3
+ \frac{343}{18} N g_2 g_3 g_4
+ \frac{686}{3} N g_2 g_3^2
+ \frac{343}{45} N g_2^2 g_4
\right. \nonumber \\
&& \left. ~~~~
+ \frac{22295}{36} N g_2^2 g_3
+ \frac{54194}{75} N g_2^3
+ \frac{1225}{36} N g_1 g_3^2
+ 196 N g_1 g_2 g_3
+ \frac{10192}{15} N g_1 g_2^2
\right. \nonumber \\
&& \left. ~~~~
+ \frac{7868}{25} N g_1^2 g_2
+ \frac{6299}{45} N g_1^3
+ \frac{343}{40} N^2 g_2^2 g_3
+ \frac{1372}{75} N^2 g_2^3
+ \frac{49}{72} N^2 g_1 g_3^2
\right. \nonumber \\
&& \left. ~~~~
+ \frac{196}{45} N^2 g_1 g_2 g_3
+ \frac{1421}{45} N^2 g_1 g_2^2
+ \frac{504}{25} N^2 g_1^2 g_2
+ \frac{853}{75} N^2 g_1^3
+ \frac{49}{180} N^3 g_1 g_2^2
\right. \nonumber \\
&& \left. ~~~~
+ \frac{7}{25} N^3 g_1^2 g_2
+ \frac{3}{10} N^3 g_1^3
+ \frac{1}{600} N^4 g_1^3
\right]
\frac{\Gamma^8 \left( \frac{2}{7} \right)
\Gamma \left( \frac{3}{7} \right) \Gamma \left( \frac{5}{7} \right)}
{\Gamma \left( \frac{6}{7} \right) \Gamma \left( \frac{4}{7} \right)}
\nonumber \\
&& +~
\left[ 
\frac{245}{9} g_3^2 g_4
+ \frac{245}{3} g_3^3
+ \frac{1372}{9} g_2 g_3^2
+ \frac{1078}{135} g_2^2 g_4
+ \frac{12446}{45} g_2^2 g_3
+ \frac{16268}{75} g_2^3
\right. \nonumber \\
&& \left. ~~~~
+ \frac{6328}{135} g_1 g_2 g_3
+ \frac{98252}{675} g_1 g_2^2
+ \frac{15128}{225} g_1^2 g_2
+ \frac{11288}{525} g_1^3
+ \frac{98}{9} N g_2 g_3^2
+ \frac{49}{45} N g_2^2 g_4
\right. \nonumber \\
&& \left. ~~~~
+ \frac{343}{15} N g_2^2 g_3
+ \frac{686}{25} N g_2^3
+ \frac{196}{27} N g_1 g_2 g_3
+ \frac{1162}{45} N g_1 g_2^2
+ \frac{1024}{75} N g_1^2 g_2
\right. \nonumber \\
&& \left. ~~~~
+ \frac{1672}{315} N g_1^3
+ \frac{49}{75} N^2 g_2^3
+ \frac{14}{45} N^2 g_1 g_2 g_3
+ \frac{77}{75} N^2 g_1 g_2^2
+ \frac{172}{225} N^2 g_1^2 g_2
\right. \nonumber \\
&& \left. ~~~~
+ \frac{212}{525} N^2 g_1^3
+ \frac{1}{75} N^3 g_1^2 g_2
+ \frac{1}{105} N^3 g_1^3
\right]
\frac{\Gamma^8 \left( \frac{2}{7} \right) \Gamma^3 \left( \frac{3}{7} \right)}
{\Gamma^3 \left( \frac{6}{7} \right)} ~+~ O(g_i^5) ~.
\end{eqnarray}
For completeness and to compare with $L_N^{\phi^8}$ we record relevant results 
for $L_N^{\phi^6}$ in our conventions which are 
\begin{eqnarray}
\gamma^{\phi^6}_\phi(g) &=& [N+2] [N+4] \frac{\pi^2}{675} g^2 ~-~ 
[N+2] [N+4] [3N+22] \frac{4\pi^3}{30375} g^3 ~+~ O(g^4) \nonumber \\ 
\beta^{\phi^6}(g) &=& -~ 4 \epsilon g ~+~ [3N + 22] \frac{4\pi}{15} g^2 
\nonumber \\
&& -~ \left[ \left[ N^3 + 34 N^2 + 620 N + 2720 \right] \pi^2 
+ 8 \left[ 53 N^2 + 858 N + 3304 \right] \right] \frac{\pi^2}{450} g^3 ~+~
O(g^4) ~. \nonumber \\
\end{eqnarray}
These are consistent with \cite{12}. One feature which is common in the 
renormalization group functions, which may be a coincidence in $L_N^{\phi^5}$ 
and $L_N^{\phi^8}$ given their potential connection through a fixed point, is 
the presence of $\Gamma \left( \frac{2}{3} \right)$. Although it occurs in the 
numerator in the $L_N^{\phi^5}$ renormalization group functions and the 
denominator of those in $L_N^{\phi^8}$ those in the latter can be replaced via
the relation
\begin{equation}
\Gamma \left( \frac{2}{3} \right) ~=~ 
\frac{2\pi}{\sqrt{3} \Gamma \left( \frac{1}{3} \right)} ~.
\end{equation}
So the renormalization group functions involve 
$\Gamma \left( \frac{1}{3} \right)$ and $\Gamma \left( \frac{2}{3} \right)$. To
explore this further we have renormalized the $O(N)$ $\phi^{12}$ theory at 
leading order which is the next candidate for a connection with an odd 
potential theory. This required computing $462$ graphs for the coupling 
constant renormalization which is an order of magnitude more than $\phi^8$ 
theory and effectively excludes 
determining the next term in the $\beta$-function. However, we find  
\begin{eqnarray}
\gamma_\phi^{\phi^{12}}(g) &=& [N + 10] [N + 8] [N + 6] [N + 4] [N + 2] 
\Gamma^{10} \left( \frac{1}{5} \right) \frac{g^2}{49792216320} ~+~ O(g^3) 
\nonumber \\
\beta^{\phi^{12}}(g) &=& -~ 10 \epsilon g ~+~ 
[5 N^3 + 750 N^2 + 19840 N + 139488 ] \Gamma^5 \left( \frac{1}{5} \right) 
\frac{g^2}{24948} ~+~ O(g^3) ~. ~~~~
\end{eqnarray}
The corresponding theory which it should have connection to is $L_N^{\phi^7}$.
This is apparent in comparing the core $\Gamma$-functions present in each set 
of renormalization group functions which are 
$\Gamma \left( \frac{1}{5} \right)$ and $\Gamma \left( \frac{2}{5} \right)$. So
at this level there appears to be a parallel connection to that of the 
$L_N^{\phi^5}$ and $L_N^{\phi^8}$ case.

While the appearance of similar $\Gamma$-functions in the renormalization group
functions of $L_N^{\phi^5}$ and $L_N^{\phi^8}$ is suggestive of a connection to
a universal theory it may not be accessible using a large $N$ approach in the 
way that the $L_N^{\phi^4}$ and $L_N^{\phi^3}$ theories were related. This is 
to do with the structure of the $\beta$-functions of each of those theories. In
particular the $N$ dependence of both $\beta$-functions follow the same 
patterns. In both theories the polynomial coefficient in $N$ of the one and two
loop terms in the $\beta$-functions are linear. This means that the critical 
coupling of the Wilson-Fisher fixed point in the $1/N$ expansion has the form
\begin{equation}
g_c ~=~ \frac{a_{11} \epsilon}{N} ~+~ 
\sum_{i=2}^\infty \left( \sum_{j=1}^\infty a_{ij} \epsilon^j \right)
\frac{1}{N^i}
\end{equation}
for each $\beta$-function where $a_{ij}$ are real numbers. In other words at
leading order there is only one term in the $\epsilon$ expansion. By contrast
if a $\beta$-function was linear in $N$ at one loop and quadratic at two loops
then the leading $1/N$ term for $g_c$ would at least be quadratic in 
$\epsilon$. This is the situation for $SU(\Nc)$ non-abelian gauge theories when
one examines $g_c$ in the large colour expansion. In fact in that case the 
degree of the polynomial in $\Nc$ at each loop order is equal to the loop 
order. This means that to find $g_c$ in a large $\Nc$ expansion in a 
non-abelian gauge theory would require the full $\beta$-function or 
equivalently the sum of an infinite number of Feynman graphs. By contrast an 
$SU(\Nc)$ non-abelian gauge theory with $\Nf$ (massless) quarks has a $1/\Nf$ 
expansion since the one and two loop terms of the $\beta$-function are linear 
in $\Nf$, \cite{49,50,51,52}. Given this property of the $\Nc$ dependence in 
the $\beta$-function of an $SU(\Nc)$ gauge theory it transpires that examining 
(\ref{rged8N}) the same feature is present for the $O(N)$ symmetry. Although 
there is a difference in that the $L_N^{\phi^8}$ $\beta$-function is quadratic 
in $N$ at one loop and quartic at two loop. However the key point is that the 
two loop term does not match the quadratic at one loop. The reason is simple to
understand from the topologies in Figures \ref{verlo8} and \ref{vernlo8} for 
example. Consequently there appears to be no critical point large $N$ expansion
method in the spirit of \cite{45,46,47} with which one could connect 
$L_N^{\phi^5}$ and $L_N^{\phi^8}$ across the dimensions and ascertain whether 
one is the ultraviolet completion of the other. Given this the $d$-dimensional 
conformal field theory formalism currently being developed in \cite{20} may 
offer the only major viable strategy to investigate this extension of the 
$L_N^{\phi^3}$ and $L_N^{\phi^4}$ connection.

\sect{Discussion.}

We have renormalized various scalar quantum field theories with odd potentials 
as well as extending these to include an $O(N)$ symmetry in this article. Our 
aim has partly been to provide independent perturbative information to 
complement other methods such as a $d$-dimensional conformal field theory 
approach where the $\epsilon$ expansion of the related critical exponents can 
be deduced. As with the renormalization group functions of scalar theories with
even order potentials the structure of the renormalization group functions does
not involve rationals at low orders. Instead combinations of $\Gamma$-functions
with rational arguments emerge. As with the widely examined $\phi^3$ and
$\phi^4$ theories the higher loop terms should introduce a new set of numbers
which should be related to the $\Gamma \left( \frac{p}{q} \right)$ where $p$
and $q$ are coprime integers. For instance in $\phi^4$ theory it is known that
the Riemann zeta series $\zeta(n)$ appears at four and higher loops. Such
numbers derive for example from coefficients in $\Gamma(n+\epsilon)$ where $n$
is an integer. Equally it is now known that multiple zeta values emerge at six 
loops after the pioneering work of \cite{40}. For the renormalization group 
functions of $\phi^r$ theories with $r$~$\geq$~$5$ a parallel numerology should
emerge. A clue to this is in the results of \cite{12} for $\phi^6$ theory where
the numbers akin to $\zeta(n)$ were extracted using the Gegenbauer polynomial
methods of \cite{53}. This technique is ideal for representing the angular 
integration in terms of nested sums. In $\phi^3$ and $\phi^4$ theory these
naturally led to $\zeta(n)$ but in \cite{12} the corresponding quantity is
Dirichlet's $\beta$-function $\beta(z)$ and specifically $\beta(2)$ and
$\beta(4)$. Given that the development of these Riemann zeta sums has led to 
the wide and systematic use of hyperlogarithms for the basis of renormalization
group functions it would seem that to tackle the next loop orders in $\phi^r$ 
theories with $r$~$\geq$~$5$ would require the development of that machinery 
by, for example, extending the {\sc Hyperint} package, \cite{54}. This will
need some care at higher loops since one will move beyond the effective 
triangle diagrams illustrated in Figures \ref{verlo5}, \ref{verlo7}, 
\ref{verlo9}, \ref{verlo8} and \ref{vernlo8}. For example, at the next level
effective boxes and pentagons with non-unit propagator exponents will arise.

While this discussion on numerology may appear disjoint the aim is to draw 
attention to it for several reasons. First, the insights deriving from 
conformal field theory methods such as \cite{20} must retain the structure of 
the renormalization group functions in its underlying algebra. Equally it must 
be informed by the structure of the Feynman diagrams in the perturbative or 
equivalently the $\epsilon$ expansion. For these higher order scalar potentials
the numbers analogous to the multiple zeta values of $\phi^4$ theory appear to 
emerge at lower loop orders. Therefore $\phi^5$ may provide the simplest 
testing ground for understanding the mathematical interconnectedness of the 
structure of non-trivial Feynman integrals further and the algebraic structure 
of the quantum field theory itself as a whole. One minor example of this was 
perhaps indicated by the extension to the $O(N)$ symmetric theories. Using an 
auxiliary field $\sigma$ a theory with an odd potential may not be unrelated to
one with an even potential in the same way that $O(N)$ $\phi^3$ theory is the 
ultraviolet completion of $O(N)$ $\phi^4$ theory. Central to the establishment 
of this was the use of the large $N$ formalism of \cite{45,46,47}. From the 
loop orders we have computed it would seem that the application of that 
particular large $N$ method may not be applicable. For it to be used one would 
have to be able to determine the location of the Wilson-Fisher fixed point at 
leading order in $1/N$. In $O(N)$ $\phi^3$ and $\phi^4$ theory this is possible
because the respective $\beta$-functions are linear in $N$ at two loops. For 
$O(N)$ $\phi^8$ theory the next-to-leading correction to that $\beta$-function 
is the same order as the leading one in terms of $1/N$. From the decoration of 
lines by bubbles due to the high order potential it would be a surprise if this
did not persist to all orders. Therefore it may be the case that the only 
realistic technique which could be used to establish any connection between 
$O(N)$ $\phi^5$ and $\phi^8$ theory as well as that between $O(N)$ $\phi^7$ and 
$\phi^{12}$ theory is that of $d$-dimensional conformal field theory. In some 
sense if these theories are connected along a thread of Wilson-Fisher fixed 
points they should have a base within the two dimensional universal theory of 
(\ref{laginftyn}). Finally, while our focus throughout has been on scalar field
theories the next suite of theories to examine in the present context of higher
order potentials are fermionic models such as the $O(N)$ Gross-Neveu model, 
\cite{55}, $O(N)$ supersymmetric nonlinear $\sigma$ models such as those 
considered in \cite{56}, or the non-abelian Thirring models, \cite{57}. For the
latter one would have to use the parallel of the auxiliary field $\sigma$ to 
effect the extension, for example, which would require higher spin fields. 

\vspace{1cm}
\noindent
{\bf Acknowledgements.} The author thanks A. Codello, D. Kreimer, P. Nogueira,
M. Safari, R.M. Simms, G.P. Vacca and O. Zanusso for discussions as well as S. 
Rychkov and S.\'{E}. Derkachov for pointing out \cite{6,7}. The diagrams were 
prepared with the {\sc Axodraw} package, \cite{58}. This work was carried out 
with the support of the STFC through the Consolidated Grant ST/L000431/1.


\begin{thebibliography}{99}
\bibitem{1} K.G. Wilson, Phys. Rev. {\bf B4} (1971), 3174.
\bibitem{2} K.G. Wilson, Phys. Rev. {\bf B4} (1971), 3184.
\bibitem{3} K.G. Wilson, Phys. Rev. Lett. {\bf 28} (1972), 548.
\bibitem{4} K.G. Wilson \& M.E. Fisher, Phys. Rev. Lett. {\bf 28} (1972), 240.
\bibitem{5} K.G. Wilson, Phys. Rept. {\bf 12} (1974), 75.
\bibitem{6} L.N. Lipatov, Sov. Phys. JETP {\bf 44} (1976), 1055.
\bibitem{7} A.B. Zamolodchikov, Sov. J. Nucl. Phys. {\bf 44} (1986), 529.
\bibitem{8} R.D. Pisarski, Phys. Rev. {\bf D28} (1983), 1554.
\bibitem{9} W.A. Bardeen, M. Moshe \& M. Bander, Phys. Rev. Lett. {\bf 52}
(1984), 1188.
\bibitem{10} R. Gudmundsdottir, G. Rydnell \& P. Salomonson, Phys. Rev. Lett.
{\bf 53} (1984), 2529.
\bibitem{11} R. Gudmundsdottir, G. Rydnell \& P. Salomonson, Annals Phys.
{\bf 162} (1985), 72.
\bibitem{12} J.S. Hager, J. Phys. {\bf A35} (2002), 2703.
\bibitem{13} J. Hofmann, Nucl. Phys. {\bf B350} (1991), 789.
\bibitem{14} J. O'Dwyer \& H. Osborn, Annals Phys. {\bf 323} (2008), 1859.
\bibitem{15} L. Zambelli \& O. Zanusso, Phys. Rev. {\bf D95} (2017), 085001.
\bibitem{16} A. Codello, N. Defenu \& G. D'Odorico, Phys. Rev. {\bf D91} 
(2015), 105003.
\bibitem{17} A. Codello, M. Safari, G.P. Vacca \& O. Zanusso, Phys. Rev. 
{\bf D96} (2017), 081701.
\bibitem{18} R. Ben Al\`{\i} Zinati \& A. Codello, J. Stat. Mech. {\bf 1801}
(2018), 013206.
\bibitem{19} F. Gliozzi, A.L. Guerrieri, A.C. Petkou \& C. Wen, JHEP {\bf 1704}
(2017), 056. 
\bibitem{20} A. Codello, M. Safari, G.P. Vacca \& O. Zanusso, JHEP {\bf 1704}
(2017), 127.
\bibitem{21} E. Br\'{e}zin, J.C. Le Guillou, J. Zinn-Justin \& B.G. Nickel,
Phys. Lett. {\bf A44} (1973), 227.
\bibitem{22} A.A. Vladimirov, D.I. Kazakov \& O.V.  Tarasov, Sov. Phys. JETP
{\bf 50} (1979), 521.
\bibitem{23} F.M. Dittes, Yu.A. Kubyshin \& O.V. Tarasov, Theor. Math. Phys.
{\bf 37} (1978), 879.
\bibitem{24} K.G. Chetyrkin, A.L. Kataev \& F.V. Tkachov, Phys. Lett. {\bf B99}
(1981), 147; {\bf B101} (1981), 457(E).
\bibitem{25} K.G. Chetyrkin, S.G. Gorishniy, S.A. Larin \& F.V. Tkachov, Phys.
Lett. {\bf B132} (1983), 351.
\bibitem{26} H. Kleinert, J. Neu, V. Schulte-Frohlinde, K.G. Chetyrkin \& S.A.
Larin, Phys. Lett. {\bf B272} (1991), 39; {\bf B319} (1993), 545(E).
\bibitem{27} D.V. Batkovich, K.G. Chetyrkin \& M.V. Kompaniets, Nucl. Phys.
{\bf B906} (2016), 147.
\bibitem{28} O. Schnetz, Phys. Rev. {\bf D97} (2018), 085018.
\bibitem{29} M.V. Kompaniets \& E. Panzer, PoS LL2016 (2016), 038.
\bibitem{30} M.V. Kompaniets \& E. Panzer, Phys. Rev. {\bf D96} (2017), 036016.
\bibitem{31} O.F. de Alcantara Bonfim, J.E. Kirkham \& A.J. McKane, J. Phys.
{\bf A13} (1980), L247; {\bf A13} (1980), 3785(E).
\bibitem{32} O.F. de Alcantara Bonfim, J.E. Kirkham \& A.J. McKane, J. Phys.
{\bf A14} (1981), 2391.
\bibitem{33} L. Fei, S. Giombi, I.R. Klebanov \& G. Tarnopolsky, Phys. Rev.
{\bf D91} (2015), 045011.
\bibitem{34} J.A. Gracey, Phys. Rev. {\bf D92} (2015), 025012.
\bibitem{35} L. Fei, S. Giombi \& I.R. Klebanov, Phys. Rev. {\bf D90} (2014),
025018.
\bibitem{36} P. Nogueira, J. Comput. Phys. {\bf 105} (1993), 279.
\bibitem{37} J.A.M. Vermaseren, math-ph/0010025.
\bibitem{38} M. Tentyukov \& J.A.M. Vermaseren, Comput. Phys. Commun. {\bf 181}
(2010), 1419.
\bibitem{39} S.A. Larin \& J.A.M. Vermaseren, Phys. Lett. {\bf B303} (1993),
334.
\bibitem{40} D.J. Broadhurst \& D. Kreimer, Int. J. Mod. Phys. {\bf C6} (1995), 
519.
\bibitem{41} D.J. Broadhurst \& D. Kreimer, Phys. Lett. {\bf B393} (1997), 
403.
\bibitem{42} F. Brown, Commun. Math. Phys. {\bf 287} (2009), 287.
\bibitem{43} O. Schnetz, Commun. Num. Theor. Phys. {\bf 4} (2010), 1.
\bibitem{44} F. Brown \& D. Kreimer, Lett. Math. Phys. {\bf 103} (2013), 933.
\bibitem{45} A.N. Vasil'ev, Y.M. Pismak \& J.R. Honkonen, Theor. Math. Phys.
{\bf 46} (1981), 104.
\bibitem{46} A.N. Vasil'ev, Y.M. Pismak \& J.R. Honkonen, Theor. Math. Phys.
{\bf 47} (1981), 465.
\bibitem{47} A.N. Vasil'ev, Y.M. Pismak \& J.R. Honkonen, Theor. Math. Phys.
{\bf 50} (1982), 127.
\bibitem{48} T. Banks \& A. Zaks, Nucl. Phys. {\bf B196} (1982), 189.
\bibitem{49} D.J. Gross \& F.J. Wilczek, Phys. Rev. Lett. {\bf 30}
(1973), 1343.
\bibitem{50} H.D. Politzer, Phys. Rev. Lett. {\bf 30} (1973), 1346.
\bibitem{51} D.R.T. Jones, Nucl. Phys. {\bf B75} (1974), 531.
\bibitem{52} W.E. Caswell, Phys. Rev. Lett. {\bf 33} (1974), 244.
\bibitem{53} K.G. Chetyrkin, A.L. Kataev \& F.V. Tkachov, Nucl. Phys. 
{\bf B174} (1980), 345.
\bibitem{54} E. Panzer, Comput. Phys. Commun. {\bf 188} (2015), 148.
\bibitem{55} D. Gross \& A. Neveu, Phys. Rev. {\bf D10} (1974), 3235.
\bibitem{56} M. Heilmann, D.F. Litim, F. Synatschke-Czerwonka \& A. Wipf, Phys.
Rev. {\bf D86} (2012), 105006.
\bibitem{57} R. Dashen \& Y. Frishman, Phys. Rev. {\bf D11} (1975), 2781.
\bibitem{58} J.C. Collins \& J.A.M. Vermaseren, arXiv:1606.01177 [cs.OH].
\end{thebibliography}
\end{document}